# The fixed-point iteration method for IMRT optimization with truncated dose deposition coefficient matrix

**Zhen Tian, Masoud Zarepisheh, Xun Jia, and Steve B. Jiang**

Center for Advanced Radiotherapy Technologies and Department of Radiation Medicine and Applied Sciences, University of California San Diego, La Jolla, CA 92093, USA

Email: sbjiang@ucsd.edu, xunjia@ucsd.edu

In the treatment plan optimization for intensity modulated radiation therapy (IMRT), dose-deposition coefficient (DDC) matrix is often pre-computed to parameterize the dose contribution to each voxel in the volume of interest from each beamlet of unit intensity. However, due to the limitation of computer memory and the requirement on computational efficiency, in practice matrix elements of small values are usually truncated, which inevitably compromises the quality of the resulting plan. A fixed-point iteration scheme has been applied in IMRT optimization to solve this problem, which has been reported to be effective and efficient based on the observations of the numerical experiments. In this paper, we aim to point out the mathematics behind this scheme and to answer the following three questions: 1) whether the fixed-point iteration algorithm converges or not? 2) when it converges, whether the fixed point solution is same as the original solution obtained with the complete DDC matrix? 3) if not the same, whether the fixed point solution is more accurate than the naive solution of the truncated problem obtained without the fixed-point iteration? To answer these questions, we first performed mathematical analysis and deductions using a simplified fluence map optimization (FMO) model. Then we conducted numerical experiments on a head-and-neck patient case using both the simplified and the original FMO model. Both our mathematical analysis and numerical experiments demonstrate that with proper DDC matrix truncation, the fixed-point iteration can converge. Even though the converged solution is not the one that we obtain with the complete DDC matrix, the fixed-point iteration scheme could significantly improve the plan’s accuracy compared with the solution to the truncated problem obtained without the fixed-point iteration.

## 1. Introduction

Treatment plan optimization is a key step in intensity modulated radiation therapy (IMRT). Advanced numerical algorithms, usually iterative ones, are widely employed to solve the optimization problem, where repeated dose calculations are conducted and a set of treatment parameters, such as beamlet intensities, are iteratively adjusted to yield a clinically acceptable dose distribution (e.g., Bortfeld, 1999; Mohan *et al.*, 1996; Spirou and Chui, 1998; Xing *et al.*, 1998; Shepard *et al.*, 2000). Therefore, fast dose calculation is indispensable for IMRT optimization to ensure that the optimal solution could be found within a reasonable amount of time.

A common way to speed up the dose calculation is to pre-calculate the so called dose-deposition coefficients (DDC) matrix (e.g., Webb and Oldham, 1996; Spirou and Chui, 1998; Jeraj and Keall, 1999; Siebers *et al.*, 2001), which parameterizes the dose contribution to each voxel from each beamlet of unit intensity and is stored as a lookup table. During the optimization process, the dose distribution can be computed by multiplying the DDC matrix with a vector consisting of the fluence map. This way, only memory access and simple matrix operations are needed, which makes the dose calculation very fast. However, a typical DDC matrix demands a large amount of computer memory for storage, and consequently slows down the optimization or even makes it impossible with a limited computer memory space under some circumstances (Thieke *et al.*, 2002; Wu *et al.*, 2003; Li *et al.*, 2011). These circumstances include cases with large target sizes, small beamlet sizes, and a large number of beams (*e.g.* in volumetric modulated arc therapy (VMAT) optimization (e.g., Otto, 2008; Men *et al.*, 2010b; Zhang *et al.*, 2010; Peng *et al.*, 2012) or in the beam orientation optimization problem (e.g., Pugachev *et al.*, 2001; Bortfeld, 2010; Jia *et al.*, 2011b). Recently, high-performance graphics processing units (GPUs) have been utilized to speed up the treatment plan optimization process due to its powerful parallel computation ability at a low cost (e.g., Men *et al.*, 2009; Men *et al.*, 2010a; Men *et al.*, 2010b; Jia *et al.*, 2011a). However, its limited memory size compared to CPU makes it even more difficult to handle a huge DDC matrix. A simple and naive way to overcome the memory limitation and/or to ensure the efficiency is to neglect those small elements of the DDC matrix. Nonetheless, this would inevitably lead to adverse impacts on the resulting plan quality (Chen *et al.*, 1995; Mohan *et al.*, 1996).

To use the truncated DDC matrix in IMRT optimization without greatly sacrificing accuracy, a hybrid scheme has been proposed (Li *et al.*, 2011). In this method, the full DDC matrix is divided into a major component and a minor component. The major one (or the truncated DDC matrix) consists of elements with values larger than a certain specified threshold. For a given voxel, these elements typically correspond to the beamlets that pass through it or its vicinity. The minor component is the remainder of the DDC matrix, which are mainly the scatter contributions of the beamlets that are far away from the voxel. The major DDC matrix is much sparser than the complete DDC matrix. The optimization process is correspondingly conducted in a two-nested-loop fashion. In the inner loop, the major DDC matrix is used as a lookup table to update the corresponding dose contribution, making this step very fast due to the greatly reduced number of non-zero elements in the major matrix. At each iteration of the inner loop, the dose to a voxel computed based on the major DDC matrix is corrected with a constant additive correction term that accounts for the dose contribution from the minor DDC component. In the outer loop, this correction term is updated using the fluence map obtained from the inner loop. In addition to the potential gain in efficiency, this method could also offer a practical solution to solve the insufficient memory problem when a huge DDC matrix is encountered. This is because only the major DDC matrix is needed to be stored in the inner loop, and in the outer loop, the dose correction term from the minor DDC matrix can be calculated via a full forward dose calculation without explicitly storing the minor DDC matrix elements. We would like to point out that, this two-loop idea has also been investigated previously for IMRT treatment plan optimization to achieve a good balance of efficiency and accuracy, where a less accurate but fast method is used for dose calculation in the inner loop while a more accurate but slow method is used to correct the dose distribution in the outer loop (e.g., Siebers *et al.*, 2002; Wu *et al.*, 2003; Wu, 2004). In this work, we will only focus on the two-loop method base on DDC matrix splitting since our main goal is to solve the memory issue for GPU-based IMRT optimization.

Despite the aforementioned advantages, the efficacy of this hybrid method is based on observations from numerical experiments (Li *et al.*, 2011). A set of mathematical questions remain unanswered regarding this scheme, including 1) does this method always converge? 2) if converges, does it converge to the solution to the original problem with the full DDC matrix? 3) if not to that solution, does it improve the quality of the resulting plan compared with the naive approach of using only the major component of the DDC matrix in optimization? This paper aims at understanding the mathematics behind this method and answering the above-mentioned questions from a mathematical perspective to a certain extent. By studying the underlying properties of a model problem, we try to shed some light on this method. We will also conduct numerical experiments to validate our mathematical analysis.

## 2. Methods and Materials

### *2.1 Optimization model*

In this paper, we employ a fluence map optimization (FMO) model (Men *et al.*, 2009). Let us decompose each beam into a set of beamlets (denoted by $J$) and the patient's CT image into a set of voxels (denoted by $I$). The FMO model can be written as:

$$x = \mathrm{argmin}_x F(Dx - T),\ x \geq 0 \qquad (1)$$

Where $x$ is a fluence map vector and $D$ is the DDC matrix with an element $D_{ij}$ representing the dose to voxel $i \in I$ from the beamlet $j \in J$ of unit intensity. In this paper, we adopt our in-house finite-size pencil beam (FSPB) dose calculation engine to obtain this DDC matrix (Gu *et al.*, 2009; Gu *et al.*, 2011). $T$ is the reference dose that is the prescription dose for PTV voxels and the threshold dose for OAR voxels. A quadratic two-sided voxel-based penalty function is used here:

$$
\begin{aligned}
&F\left(\sum_j D_{ij}x_j - T_i\right) = F^+\left(\sum_j D_{ij}x_j - T_i\right) + F^-\left(\sum_j D_{ij}x_j - T_i\right),\\
&F^+\left(\sum_j D_{ij}x_j - T_i\right) = \omega_i^+ \max\left\{0, \sum_j D_{ij}x_j - T_i\right\}^2, \quad i \in I, \qquad (2)\\
&F^-\left(\sum_j D_{ij}x_j - T_i\right) = \omega_i^- \max\left\{0, T_i - \sum_j D_{ij}x_j\right\}^2, \ i \in I_{PTV}.
\end{aligned}
$$

where $\omega_i^+$ and $\omega_i^-$ denote the overdosing and the underdosing penalty weights for the voxel $i$, respectively.

*2.2 Fixed-point iteration scheme*

We split the matrix $D$ into two components, $D_1$ and $D_2$ according to a user-defined threshold. $D_1$, the major component (also called the truncated matrix $D$), consists of elements of the matrix $D$ with values higher than the threshold. For a given voxel, these elements typically correspond to the beamlets that pass through it or its vicinity. $D_2$, the minor component, represents the scatter dose to a voxel from beamlets passing at large distances, usually corresponding to the long tail of the beamlet dose profile. A hybrid dose correction method (Li *et al.*, 2011) can be introduced into our FMO model as follows:

$$x^{(k+1)} = \text{argmin}_x F\left(D_1 x + \delta^{(k)} - T\right), \qquad (3)$$

$$\delta^{(k+1)} = D_2 x^{(k+1)}. \qquad (4)$$

This method consists of two loops of iterations. $k$ denotes the iteration index of the outer loop. Eq. (3), representing the inner loop, is solved using an iterative algorithm for a constant value of $\delta^{(k)}$. $\delta^{(k)}$ is the part of dose corresponding to the minor component $D_2$. Note that only the major component $D_1$ is needed for the inner loop. Due to the much reduced number of non-zero elements in $D_1$ in comparison with the full matrix $D$, the inner loop can converge much quicker than the original optimization problem with the full matrix $D$. The outer loop, represented by Eq. (4), updates the correction term $\delta^{(k+1)}$ using the minor component $D_2$ and the updated beamlet intensities $x^{(k+1)}$. We would like to point out that $D_2$ does not need to be pre-computed and stored since Eq. (4) can be computed using a forward full dose calculation in combination with the pre-computed $D_1$. Although each forward full dose calculation takes some time and needs to be done at each iteration of the outer loop, the total number of outer loop iterations is small and thus the added time is not a big concern. In this procedure, we only need to allocate computer memory for $D_1$ and $\delta$, both of which are much smaller than $D$ , and thus the memory deficiency issue is overcome.

By plugging Eq. (4) into Eq. (3), this hybrid dose correction scheme can be written as

$$x^{(k+1)} = \text{argmin}_x F\left(D_1 x + D_2 x^{(k)} - T\right) \equiv G\left(x^{(k)}\right). \qquad (5)$$

Here, $G(\ )$ is a function that denotes the mapping from $x^{(k)}$ to $x^{(k+1)}$. Such an iterative scheme that gives rise to a sequence $x^{(0)}, x^{(1)}, x^{(2)}, \ldots$, is called fixed-point iteration (FPI) in mathematics literature (Burden and Faires, 2011). Provided that $G(x)$ is continuous and the algorithm converges, this sequence approaches to a point that is called fixed point defined as $G(x) = x$. According to the Banach fixed point theorem (Palais, 2007), if there exists a positive constant $K$ such that $\frac{|G(x)-G(y)|}{|x-y|} < K < 1$ for all $x$ and $y$ in the domain of $G$, the iteration converges to its unique fixed point.

*2.3 Analysis of a simplified model*

*2.3.1 A simple quadratic model*

It is not easy to explicitly derive a closed form expression of $G$ in the existence of non-negativity constraints and the different penalty weights for underdosing and overdosing in the FMO model. To make the analysis possible, let us simplify the quadratic model. Our simple quadratic model is as follows:

$$F\left(D_1 x + D_2 x^{(k)} - T\right) = \left\|\sqrt{\lambda} D_1 x + \sqrt{\lambda} D_2 x^{(k)} - \sqrt{\lambda} T\right\|_2^2, \tag{6}$$

$$x^{(k+1)} = \operatorname{argmin}_x \left\|\sqrt{\lambda} D_1 x + \sqrt{\lambda} D_2 x^{(k)} - \sqrt{\lambda} T\right\|_2^2 = G\left(x^{(k)}\right), \tag{7}$$

where $\lambda$ is a diagonal matrix whose dimension is equal to the number of voxels. Its entries correspond to the penalty weights for each voxel. We will also ignore the non-negativity constraints on the solution for the sake of simplicity. The validity of this simple model for the original problem will be discussed wherever is applicable.

*2.3.2 Convergence of FPI*

The optimality condition of the least-square problem in Eq. (7) is:

$$2\left(\sqrt{\lambda} D_1\right)^T \left(\sqrt{\lambda} D_1 x + \sqrt{\lambda} D_2 x^{(k)} - \sqrt{\lambda} T\right) = 0. \tag{8}$$

Assume $D_1$ is full column rank and all the diagonal elements of $\lambda$ are nonzero, $D_1^T \lambda D_1$ is full rank and hence invertible. It can be readily shown that:

$$G\left(x^{(k)}\right) = x^{(k+1)} = (D_1^T \lambda D_1)^{-1}(D_1^T \lambda T - D_1^T \lambda D_2 x^{(k)}), \tag{9}$$

$$\frac{|G(x)-G(y)|}{|x-y|} = |-(D_1^T \lambda D_1)^{-1} D_1^T \lambda D_2|. \tag{10}$$

From Eq. (10), we can see that whether the FPI converges or not is completely dependent on the magnitude of the term $|-(D_1^T \lambda D_1)^{-1} D_1^T \lambda D_2|$. In essence, it depends on how we truncate the DDC matrix $D$. Once the matrix $D_2$ is relatively small so that the condition $\frac{|G(x)-G(y)|}{|x-y|} < K < 1$ is satisfied, the FPI in Eq. (7) converges to the unique fixed point of the function $G\left(x^{(k)}\right)$. It is worthwhile to mention that this is a sufficient, but not the necessary, condition to ensure the convergence.

Despite the simplification that we made, we expect that the main conclusions drawn here are still qualitatively valid due to the captured quadratic nature of the problem. Therefore, although we cannot get the close form of function $G(.)$ for the original FMO model, it is expected that there should exist a truncation threshold to make $D_2$ relatively small, so that the condition $\frac{|G(x)-G(y)|}{|x-y|} < K < 1$ is satisfied and FPI converges. However, we do not expect that this truncation threshold for the full FMO model is the same as that for the simplified model.

### *2.3.3 Fixed point solution*

After analyzing the convergence of FPI, we come to the second question, namely, given $D_1$ and $D_2$ that make FPI converge, whether the obtained fixed point solution $x^*$ is the solution $x^o$ to the original optimization problem with the full DDC matrix $D$, namely

$$x^o = \text{argmin}_x \left\| \sqrt{\lambda} D x - \sqrt{\lambda} T \right\|_2^2 = (D^T \lambda D)^{-1} D^T \lambda T. \tag{11}$$

Assuming that FPI converges to its fixed point, *i.e.* $x^{(k)} \to x^*$, we can take the limit of $k \to \infty$, yielding

$$x^* = (D_1^T \lambda D)^{-1} D_1^T \lambda T. \tag{12}$$

Unlike what we hoped, it turns out that even though FPI converges with a proper truncation of the DDC matrix $D$, the obtained solution $x^*$ is not the solution $x^o$ to the original optimization problem we would like to solve. A general proof demonstrating this phenomena still holds in the original problem is given in Appendix A.

Given this fact, an important question would be whether the FPI scheme improves the resulting plan's accuracy, compared with the solution $x^t$ to the naive optimization problem using only the truncated DDC matrix $D_1$, namely

$$x^t = \text{argmin}_x \left\| \sqrt{\lambda} D_1 x - \sqrt{\lambda} T \right\|_2^2 = (D_1^T \lambda D_1)^{-1} D_1^T \lambda T. \tag{13}$$

Since it is meaningful and of clinical interest to justify the plan quality in terms of dose distribution, we would like to compare $\left\| \sqrt{\lambda} D x^* - \sqrt{\lambda} D x^o \right\|_2$ and $\left\| \sqrt{\lambda} D x^t - \sqrt{\lambda} D x^o \right\|_2$ instead of $\|x^* - x^o\|_2$ and $\|x^t - x^o\|_2$. This is essentially a quantity regarding the difference of the resulting dose distribution weighted by the voxel importance parameters used in the optimization problem $\lambda$. Note that according to the derivation shown in Appendix B, it can be concluded that, as long as the minor component $D_2$ is small enough such that its high order term could be neglected, the fixed point solution $x^*$ would be closer to the original $x^o$ than the naive solution $x^t$ to $x^o$ in dose domain.

## *2.4 Numerical experiments*

To conform our mathematical analysis numerically, we choose a head-and-neck (H&N) patient case. There are 6 beams in the treatment plan, with 5 coplanar beams at $0^\circ$ couch angle and 1 noncoplanar beam at $90^\circ$ couch angle. The source-to-axis distance is 100 cm. The number of voxels used in optimization is 25744, with a voxel size of $0.273\times0.273\times0.125$ cm$^3$. We only consider the beamlets that are inside the beam-eye-view projection of the PTV at each beam angle, so that all the beamlets would have primary dose contributions to at least one voxel. Moreover, since the dose contributions of two beamlets to the voxels could not be exactly the same due to their different locations, theoretically $D$ and $D_1$ should consist of linearly independent columns and thus be full column rank. In this paper, we adopt two different beamlet sizes, namely $1\times1$ cm$^2$ and $0.5\times0.2$ cm$^2$ for the experiments, and the corresponding numbers of beamlets are 444 and 4788, respectively.

In our experiments, the $D$ matrix is pre-calculated using our in-house FSPB dose calculation algorithm (Gu *et al.*, 2009). It is further divided into the major component $D_1$ and the minor component $D_2$ using a truncation threshold $\gamma$. Specifically, for each beamlet, we set a circular cone centered at the beamlet central axis with the cone radius at isocenter level being $\gamma$. In the matrix $D$, those elements corresponding to the voxels inside this cone belong to the major component $D_1$ and others belong to $D_2$. Here, we assume that the complete matrix $D$ is obtained with $\gamma$ equal to 10 cm. Then the smaller $\gamma$ is, the more aggressive the truncation is.

In this paper, our FMO model with the FPI scheme is implemented in Compute Unified Device Architecture (CUDA) programming environment and using an NVIDIA Tesla C1060 GPU card to speed up the calculation. Note that for the simplified quadratic model, since we have already derived the closed form expression of the solutions in which only matrix operations are included, MatLab codes are used to perform corresponding calculations.

## 3. Experimental Results

### *3.1 Results for the simplified quadratic model*

First, numerical experiments with the simplified quadratic optimization model in Eq. (7) are used to verify our mathematical deduction. Because of the matrix operations involved in Eq. (10~13), we use the bigger beamlet size (1×1cm$^2$) to yield relatively small matrices, so that those matrix operations can be handled in MatLab accurately. The convergence of FPI is tested with this simplified model at different levels of truncation for the matrix $D$. If FPI converges at a certain level of truncation, we calculate the fixed point solution $x^*$ and the naive solution $x^t$ obtained directly using Eq. (12) and Eq. (13). In order to evaluate the accuracy, their relative errors to the original solution $x^o$ in dose domain, denoted as $e^*$ and $e^t$ respectively, are calculated as follows:

$$e^* = \frac{\left\|\sqrt{\lambda}Dx^* - \sqrt{\lambda}Dx^o\right\|_2}{\left\|\sqrt{\lambda}Dx^o\right\|_2} \tag{14}$$

$$e^t = \frac{\left\|\sqrt{\lambda}Dx^t - \sqrt{\lambda}Dx^o\right\|_2}{\left\|\sqrt{\lambda}Dx^o\right\|_2} \tag{15}$$

**Table 1**. Relative errors of the resulting solutions on the simplified model. $e^*$ is not available in the last case due to the non-convergence of FPI.

| $\gamma$ (cm) | $\frac{\lvert G(x)-G(y)\rvert}{\lvert x-y\rvert}$ | $e^*$ | $e^t$ |
|---|---|---|---|
| 9 | 0.0068 | 5.18e-5 | 3.12e-4 |
| 8 | 0.0161 | 1.10e-4 | 1.10e-3 |
| 7 | 0.0277 | 1.95e-4 | 3.30e-3 |
| 6 | 0.0426 | 2.56e-5 | 7.64e-3 |
| 5 | 0.0647 | 3.21e-4 | 1.49e-2 |
| 4 | 0.0972 | 4.76e-4 | 2.66e-2 |
| 3 | 0.1980 | 8.66e-4 | 4.55e-2 |
| 2 | 0.6174 | 2.43e-3 | 7.39e-2 |
| 1 | 1.2856 | 9.33e-3 | 1.60e-1 |
| 0.5 | 5.4275 | NA | 9.15e-1 |

The results are listed in Table 1. We would like to point out that, when the truncation threshold $\gamma$ is decreased, the cut-off value for including elements into $D_1$ is increased and the number of elements in $D_1$is decreased. As a consequence, the solution $x^t$ obtained by only considering the major component $D_1$ in the optimization becomes less accurate, indicated by the increased relative error when compared to the original solution $x^o$ in dose domain. In contrast, with the FPI scheme, as long as $D_2$ is relatively small, such that the expression $\frac{|G(x)-G(y)|}{|x-y|}$ is smaller than 1, this scheme could converge to its unique fixed-point solution $x^*$, which has a much smaller relative error. This observation verifies our mathematical analysis and demonstrates that FPI can take the minor component $D_2$ into account during the optimization and thus improve the accuracy of the resulting plan.

Figure 1 shows the relative error of the solution $x^{(k)}$ obtained during FPI (from Eq. (9)) in dose domain, with the truncation threshold of 2 cm, 1 cm and 0.5 cm, respectively. Here, we use 0 as the initial value for FPI, *e.g.* $x^{(0)} = 0$. We can see that when the truncation threshold is 2 cm, the solution becomes closer and closer to the ideal solution with the iterations and then fixed at a certain point, which illustrates that although FPI does not converge to the exact ideal solution, it approaches to the vicinity of the ideal solution. Moreover, it is observed that only a few iterations are required to obtain a good solution. With the 1 cm truncation threshold, the corresponding $\frac{|G(x)-G(y)|}{|x-y|}$ is slightly bigger than 1, and the relative error of the solution to the ideal solution decreases at the first 5 iterations. After that it begins to increase very slowly, which is difficult to be observed in Figure 1. Due to this extreme slow increase, we still calculated the relative error of the solution obtained at the 50$^{\text{th}}$ iteration and show it in Table 1. When the truncation threshold is decreased further to be 0.5 cm, the corresponding $\frac{|G(x)-G(y)|}{|x-y|}$ becomes 5.4275. Consequently, after the first 26 iterations of decreasing, the relative error oscillates at the next 10 iterations and then keeps increasing, shown in Figure 1.

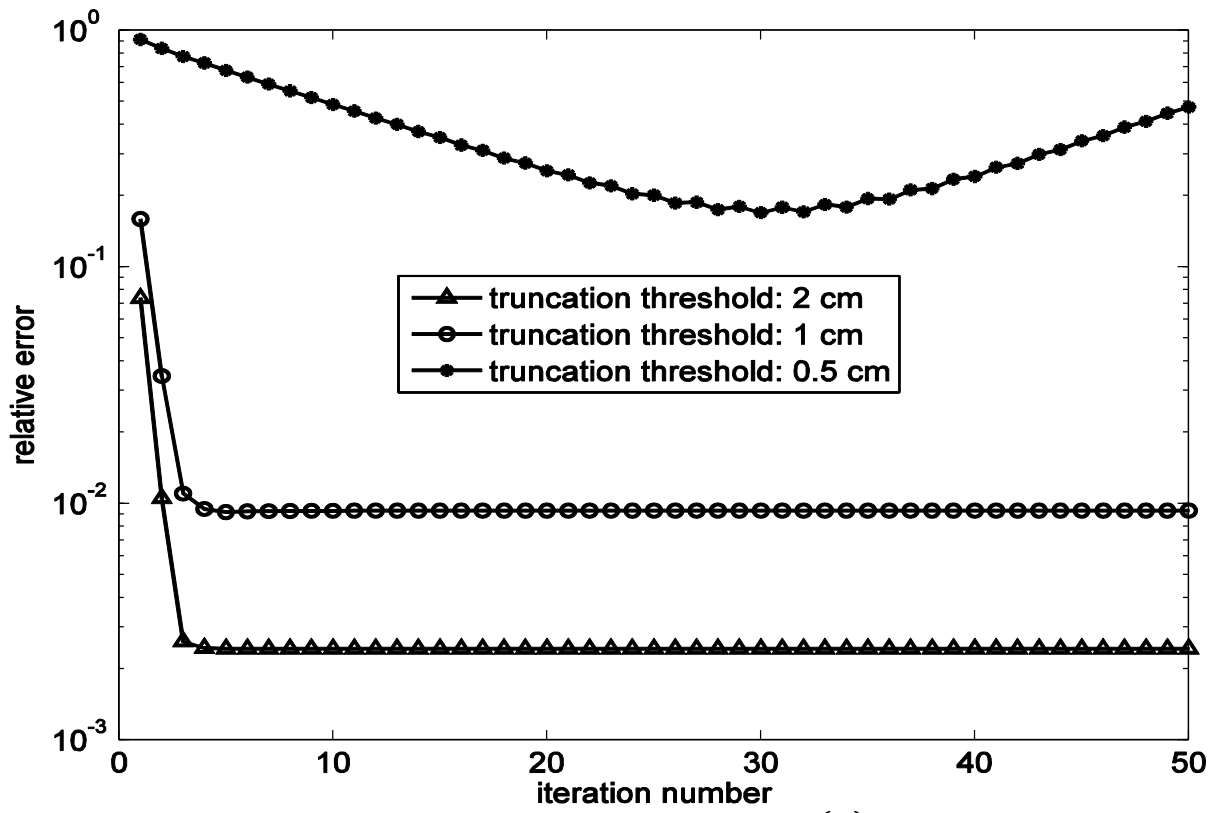

**Figure 1.** The relative error of the solution $x^{(k)}$ obtained during FPI when compared to the original solution $x^o$ in dose domain.

*3.2 Results for the FMO model*

We further test the FPI scheme on the FMO model stated in Eq. (5). At first, we still adopt the bigger beamlet size (1×1cm$^2$) and calculate the relative error of the solution obtained with FPI and that of the naive solution obtained without FPI, both with respect to the ideal solution. The experimental results are listed in Table 2. Since Figure 1 shows that only a few iterations are required to get a good result, here we only employ three FPI iterations for all different levels of truncations. Note that when the truncation threshold $\gamma$ is 10 cm, the truncation is very little, so the matrix $D_1$ in this case is assumed to be the complete DDC matrix $D$ and thus the solution $x^o$ obtained is actually the solution to the original problem. Apparently, with the truncation threshold being decreased, the number of nonzero elements in the major matrix $D_1$ becomes smaller, thus the computation time without FPI is shortened. However, the gradually increased relative error of $x^t$ reveals that the accuracy is also sacrificed to a larger degree due to the truncation. We can see from Table 2 that for all truncation levels, the solutions obtained with three FPI iterations have much smaller relative errors than the naive solutions obtained without using FPI, which demonstrates again that FPI can take the DDC matrix truncation into consideration during optimization and thus improve the plan accuracy. Note that since our algorithm is implemented on GPU to realize parallel computation, the improvement of efficiency may not be as conspicuous as one may expect. For GPU applications, the memory limitation is our main concern and the major motivation in this work is to solve the memory deficient issue with this FPI scheme.

**Table 2**. Experimental results with the FMO model. $N$ represents the number of the nonzero elements in the corresponding major matrix $D_1$. $T^*$ and $T^t$ denote the running time on GPU to get the solution $x^{(k)}$ with three FPI iterations and the time to get the naive solution $x^t$ without FPI, respectively. $e^*$ and $e^t$ denote the relative error of the solution obtained with FPI and that of the naive solution without FPI, respect to the ideal solution. $T^*$and $e^*$ are not available when the truncation threshold $\gamma$ equals 10 cm, since there is no truncation in this case and the ground truth solution is obtained.

| $\gamma$ (cm) | $N$ | $T^*$ (sec) | $T^t$ (sec) | $e^*$ | $e^t$ |
|---|---|---|---|---|---|
| 10 | 10747994 | NA | 2.62 | NA | 0 |
| 8 | 10093295 | 4.85 | 1.80 | 1.90e-7 | 2.30e-6 |
| 6 | 8356153 | 4.02 | 1.60 | 7.24e-7 | 7.44e-5 |
| 4 | 5049276 | 2.93 | 0.66 | 9.75e-7 | 7.38e-4 |
| 2 | 1525861 | 2.04 | 0.77 | 1.13e-4 | 5.29e-3 |
| 1 | 393023 | 1.91 | 0.42 | 1.22e-3 | 2.43e-2 |

To further investigate the plan quality improvement using this FPI scheme, we also adopt a clinically realistic beamlet size (0.5×0.2 cm$^2$) and still adopt three FPI iterations. With the truncation threshold $\gamma$ chosen to be 2 cm, the dose-volume histograms (DVHs) of the resulting plans obtained without or with FPI are shown in Figures 2(a) and 2(b), respectively. Since the naive plan $x^t$ is obtained by only considering $D_1$ during the IMRT optimization, the dotted DVH curves of the optimized dose $D_1x^t$ are very close to those of the original plan's dose distribution $Dx^0$, shown in Figure 2(a), which illustrates that the optimization is carried out very successfully. However, after adding the dose contribution $D_2x^t$ that was ignored in the optimization process, the actual dose $Dx^t$ that is delivered to the patient is higher. In particular, the PTV is overdosed by about 5.8%. In contrast, by applying FPI to the IMRT optimization, both the major and the minor part of the matrix $D$ are considered in the FMO model. Therefore, the dotted DVH curves of the major dose part $D_1x^*$ appears lower than those of the original plan's dose distribution $Dx^0$, shown in Figure 2(b). However, after adding the minor dose portion, the DVHs of the actual dose $Dx^*$ match those of $Dx^0$ very well. Similar results are also found for DVHs of optic nerves, which are not shown here for the sake of clarity.

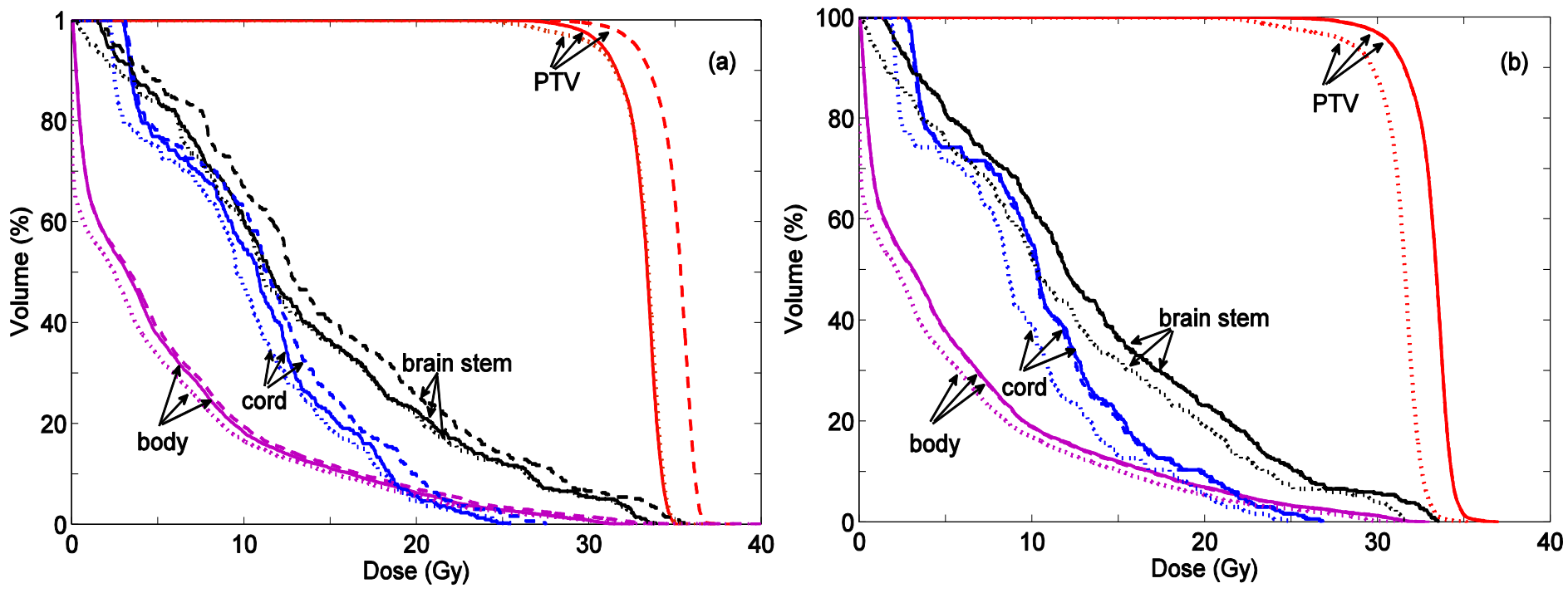

**Figure 2.** Dose-volume histograms. (a) The naive plan $x^t$ obtained using the truncated DDC matrix $D_1$ without FPI. Solid line depicts the original plan's dose distribution $Dx^0$; dotted line represents the optimized major dose $D_1x^t$; dashed line illustrates the actual dose $Dx^t$. (b) The resulting plan $Dx^*$ obtained with three FPI iterations. Solid line denotes the original plan's dose distribution $Dx^0$; dotted line represents the optimized

major dose $D_1 x^*$; dashed line depicts the actual dose $Dx^*$.

The actual dose distributions of these plans are shown in Figure 3. The first row shows the dose distribution of the plan $Dx^0$. We did not show the dose distributions of $\mathrm{Dx}^{\mathrm{t}}$ and $\mathrm{Dx}^*$, as visually they are indistinguishable from $Dx^0$. Instead, the absolute difference between $Dx^t$ and $Dx^o$, and that between $Dx^*$ and $Dx^o$ are shown in the last two rows, respectively, with a different color map scale for a better visualization. It is apparent from these pictures that for the naive plan $x^t$ obtained by only considering the major dose contribution $D_1 x^t$ in the IMRT optimization, its actual dose $Dx^t$ is much higher compared with the original plan's dose distribution $Dx^o$, particularly for PTV. In contrast, the dose distribution of the plan obtained with three FPI iterations, $Dx^*$, looks very similar to that of the original plan and could be regarded as clinically equivalent.

## 4. Discussion and Conclusions

To solve the efficiency and/or memory issue caused by the huge DDC matrix in an IMRT optimization problem, a dose-correction scheme has been previously proposed and reported to be effective (Li *et al.*, 2011). In this paper, we have conducted some mathematical investigations to reveal the properties of the method and answer some important questions. This dose-correction method belongs to the so-called FPI scheme, which has been widely developed in mathematics literature and we can take advantage of the existing results. According to the Banach fixed point theorem, the convergence of this FPI scheme can be analyzed based on the rate of the changes of the fixed point function. To make the mathematical deduction easier, we first considered a simplified quadratic optimization model. The analysis conducted on this model can be extended to the original FMO problem and the conclusions still hold to a certain extent. The properties of this FPI scheme are also verified with numerical experiments.

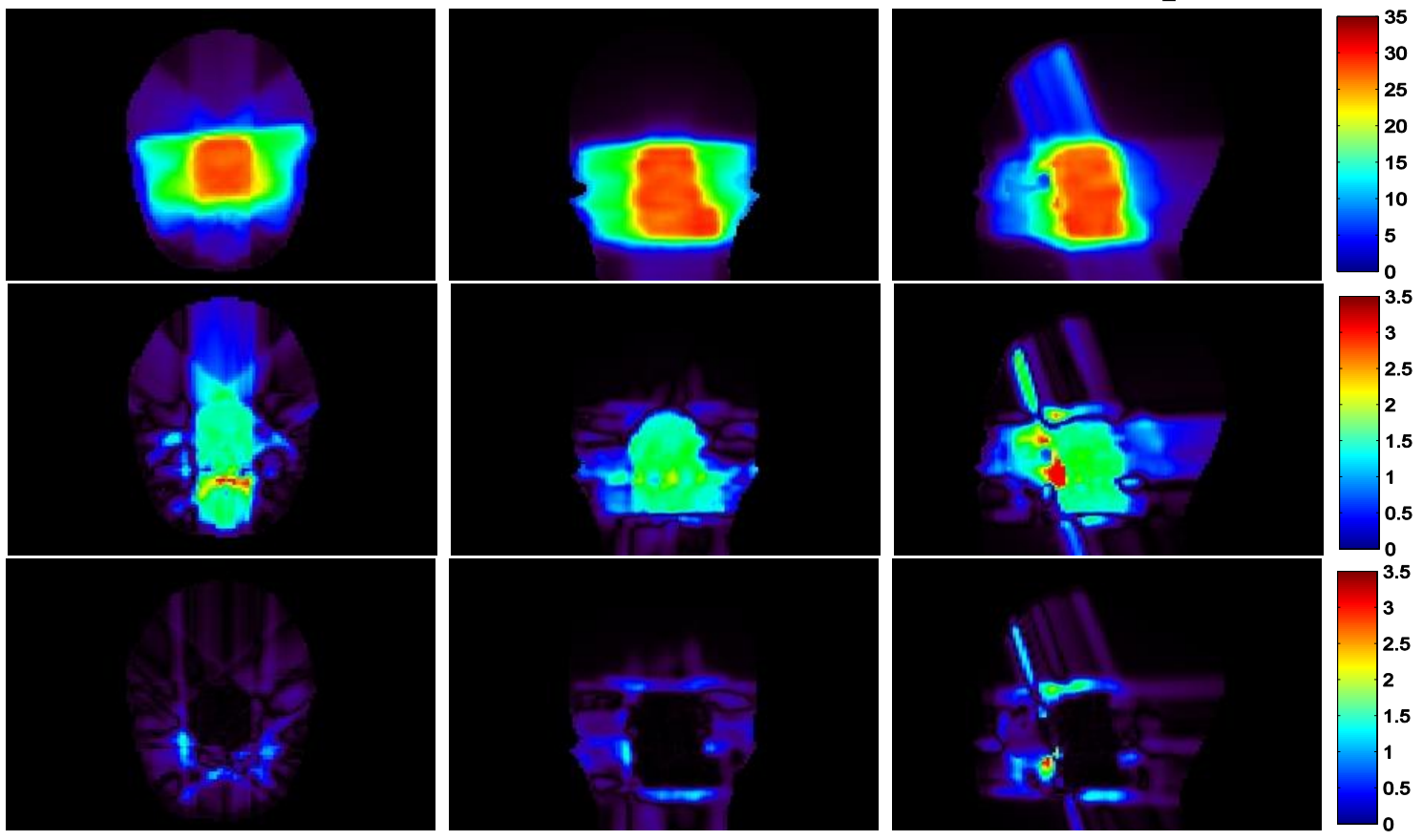

**Figure 3.** Dose distribution of the resulting plans. The first row shows the dose distribution of the plan $Dx^0$. The last two rows illustrate the absolute difference between $Dx^0$ and $Dx^t$ and that between $Dx^0$ and $Dx^*$. The three columns depict the transverse, coronal and sagittal views, respectively.

We first demonstrated that the convergence behavior of the FPI algorithm heavily depends on how we truncate the DDC matrix. In a mathematically inaccurate language, Eq. (10) indicates that the convergence of this scheme is determined by the quotient of the matrix $D_2$ and $D_1$. This is consistent with our intuition that, only when the truncation is not too heavy so that $D_1$ is still large enough to capture all the primary dose contribution, could the inner loop that updates the beamlet intensity according to $D_1$ provide a good approximation and thus make the FPI scheme work.

After examining the convergence behavior, we came to the second question that is: when it converges to $x^*$, whether this would be the ideal solution $x^o$ obtained using the complete DDC matrix. Yet, our analysis showed that these two solutions are not necessarily the same. Given this fact, a natural question that arises here is whether our fixed point solution $x^*$ is closer to the ideal solution $x^o$ than the naive solution $x^t$ to $x^o$. For the simple quadratic problem, apparently from Eq. (11~13), the solution $x^*$ is closer to $x^o$ than $x^t$, and our calculations also showed that the dose distribution corresponding to $x^*$ is a better approximation to the true dose distribution corresponding to $x^o$. We have numerically demonstrated this point in both the simplified quadratic problem and the original FMO problem.

It may be claimed that, in practice, after optimization, by renormalizing the actual dose according to the DVH curves of the PTVs, the accuracy problem due to the truncated DDC matrix can be alleviated. However, the disadvantage of this normalization method is that since the dose contribution corresponding to the matrix $D_2$ varies from voxel to voxel, uniform normalization may not yield an optimal solution, particularly for the organs at risk.

At the end of this paper, we would also like to point out that this FPI scheme may have a broader applications in IMRT optimization problems, in addition to handle the DDC matrix truncation and memory issues. For instance, when the final forward dose engine (e.g., Monte Carlo simulation) is different from the dose engine used to compute the DDC matrix in optimization (e.g., a pencil-beam model) (Siebers *et al.*, 2001; Siebers *et al.*, 2002), FPI may be employed. Another example is when some practical issues cannot be easily handled in the optimization, such as MLC transmission (e.g., LoSasso *et al.*, 1998; Seco *et al.*, 2001; Yang and Xing, 2003), FPI may help to effectively take these issues into consideration.

**Acknowledgements**

This work is supported in part by the University of California Lab Fees Research Program.

**Appendix**

*Appendix A*

Since $x^o$ is an optimal solution of the problem in Eq. (1) and $D = D_1 + D_2$, we have

$$F(D_1x^o + D_2x^o - T) \leq F(D_1x + D_2x - T) \qquad \forall x \geq 0 \tag{A1}$$

Given that $x^*$ is an optimal solution of the problem in Eq. (5), we have

$$F(D_1x^* + D_2x^* - T) \le F(D_1x + D_2x^* - T) \qquad \forall x \ge 0 \tag{A2}$$

Comparing (A1) and (A2) makes it clear that $x^*$ and $x^o$ are not necessarily the same.

*Appendix B*

We would like to compare $\left\|\sqrt{\lambda}Dx^* - \sqrt{\lambda}Dx^o\right\|_2$ and $\left\|\sqrt{\lambda}Dx^t - \sqrt{\lambda}Dx^o\right\|_2$. First, to simplify notation, we use $D_\lambda$ and $D_{1\lambda}$ to denote $\sqrt{\lambda}D$ and $\sqrt{\lambda}D_1$, and rewrite the variables as

$$\sqrt{\lambda}Dx^o = D_\lambda\left(D_\lambda^T D_\lambda\right)^{-1} D_\lambda^T \sqrt{\lambda}T \equiv P^o\sqrt{\lambda}T \tag{B1}$$

$$\sqrt{\lambda}Dx^* = D_\lambda\left(D_{1\lambda}^T D_\lambda\right)^{-1} D_{1\lambda}^T \sqrt{\lambda}T \equiv P^*\sqrt{\lambda}T \tag{B2}$$

$$\sqrt{\lambda}Dx^t = D_\lambda\left(D_{1\lambda}^T D_{1\lambda}\right)^{-1} D_{1\lambda}^T \sqrt{\lambda}T \equiv P^t\sqrt{\lambda}T \tag{B3}$$

Since $D = D_1 + D_2, D_2 \ll D_1$, we can consider $D_2$ as a small perturbation and denote $D_{1\lambda}$ as $D_\lambda - E$ and rewrite $P^*$ and $P^t$ as follows:

$$\begin{aligned} P^* &= D_\lambda\left(\left(D_\lambda^T - E^T\right)D_\lambda\right)^{-1}\left(D_\lambda^T - E^T\right) \\ &\approx P^o + D_\lambda\left(D_\lambda^T D_\lambda\right)^{-1} E^T P^o - D_\lambda\left(D_\lambda^T D_\lambda\right)^{-1} E^T + o(E^2), \end{aligned} \tag{B4}$$

$$\begin{aligned} P^t &= D_\lambda\left(\left(D_\lambda^T - E^T\right)\left(D_\lambda - E\right)\right)^{-1}\left(D_\lambda^T - E^T\right) \\ &\approx P^o + P^o E\left(D_\lambda^T D_\lambda\right)^{-1} D_\lambda^T + \left(D_\lambda^T D_\lambda\right)^{-1} E^T P^o - D_\lambda\left(D_\lambda^T D_\lambda\right)^{-1} E^T + o(E^2). \end{aligned} \tag{B5}$$

Since $(P^o)^2 = P^o$, $(P^o)^T = P^o$, $P^o D_\lambda = D_\lambda$, and $D_\lambda^T P^o = D_\lambda^T$, it follows

$$\begin{aligned} &(P^t - P^o)^T(P^t - P^o) - (P^* - P^o)^T(P^* - P^o) \\ &\approx -E\left(D_\lambda^T D_\lambda\right)^{-1} D_\lambda^T E\left(D_\lambda^T D_\lambda\right)^{-1} D_\lambda^T - D_\lambda\left(D_\lambda^T D_\lambda\right)^{-1} E^T D_\lambda\left(D_\lambda^T D_\lambda\right)^{-1} E^T \\ &+ D_\lambda\left(D_\lambda^T D_\lambda\right)^{-1} E^T P^o E\left(D_\lambda^T D_\lambda\right)^{-1} D_\lambda^T + D_\lambda\left(D_\lambda^T D_\lambda\right)^{-1} E^T D_\lambda\left(D_\lambda^T D_\lambda\right)^{-1} E^T P^o \\ &+ P^o E\left(D_\lambda^T D_\lambda\right)^{-1} D_\lambda^T E\left(D_\lambda^T D_\lambda\right)^{-1} D_\lambda^T, \end{aligned} \tag{B6}$$

where $H = D_\lambda\left(D_\lambda^T D_\lambda\right)^{-1} E^T$.

Since $HP^oH = HH, H^TP^oH^T = H^TH^T$, then

$$\begin{aligned} &\|P^t - P^o\|_2^2 - \|P^* - P^o\|_2^2 \\ &= tr(-H^TH^T - HH + HP^oH^T + HHP^o + P^oH^TH^T) \\ &= tr(-H^TH^T - HH + HP^oH^T + HP^oH + H^TP^oH^T) \\ &= tr(HP^oH^T) = tr(HP^o(HP^o)^T) \ge 0 \end{aligned} \tag{B7}$$